\documentclass[a4paper,11pt]{article}
\usepackage{pos}
\usepackage{amsmath}
\usepackage{mathtools}
\usepackage{subcaption}
\usepackage{graphicx}
\usepackage{placeins}
\usepackage{mathtools}
\usepackage{float}

\title{Non-perturbative renormalization of the energy momentum tensor in the 2d O(3) nonlinear sigma model}
\ShortTitle{Non-perturbative renormalization of the EMT in the 2d O(3) nlsm}

\author*[a]{Mika Lauk}
\author[a,b]{Agostino Patella}

\affiliation[a]{Humboldt Universit\"at zu Berlin, Institut f\"ur Physik, \\ Zum Grossen Windkanal 6, 12489 Berlin, Germany}
\affiliation[b]{DESY, Platanenallee 6, D-15738 Zeuthen, Germany}

\emailAdd{mika.akim.lauk@hu-berlin.de}

\abstract{The two-dimensional O(3) nonlinear sigma model is a well known toy model for studying non-perturbative phenomena in quantum field theory. A central challenge is the renormalization of the energy-momentum tensor, which is complicated by the nonlinear realization of the $O(3)$ symmetry leading to non-trivial operator mixing patterns, and by large discretization artifacts affecting the determination of renormalization constants. We present results for the renormalization constants in the non-singlet sector, employing a modified lattice action with shifted boundary conditions and defining the renormalized coupling through the gradient flow. With this we obtain a precise determination of the renormalization constants $z_T$ and $Z_T$.}

\FullConference{The 42nd International Symposium on Lattice Field Theory (LATTICE2025)\\
2-8 November 2025\\
Tata Institute of Fundamental Research, Mumbai, India\\}

\begin{document}
\maketitle
\section{Introduction}
The two dimensional $O(3)$ nonlinear sigma model is an interesting quantum field theory that is relatively simple to define yet shares many features with Yang-Mills theory and QCD, such as asymptotic freedom, a non-trivial topological structure and a dynamically generated mass gap. These shared features make it a popular testbed for QCD, especially in the development of methods in lattice field theory. Additionally, the target space $O(3)$ can be extended supersymmetrically \cite{Costa:2022ezw}, providing a playground to investigate aspects of strings in $AdS$ whose worldsheet discretization remains a challenge \cite{Bliard:2022oof}.

An interesting starting point in dealing with this model is the non-perturbative definition of the energy momentum tensor (EMT). As the lattice discretization breaks translational symmetry,
the EMT is no longer conserved and has to be renormalized \cite{Caracciolo:1988hc}. This is a non-trivial task as the $O(3)$ symmetry is nonlinearly realized, leading to non-trivial operator mixing \cite{Brezin:1976ap}. Additionally, the theory exhibits large discretization artifacts \cite{Balog:2009np} that complicate extraction of renormalization constants.

Focusing on the non-singlet sector, we determine the renormalization constants of the EMT using shifted boundary conditions \cite{Giusti:2012yj} in a finite-volume gradient flow scheme. To reduce discretization artifacts, we employ a modified lattice action, similar to the one used in \cite{Balog:2012db}.
\subsection{The 2d $O(3)$ sigma model and its energy-momentum-tensor}

The $O(3)$ nonlinear sigma model is defined through a scalar field $\phi$ mapping a flat two-dimensional spacetime into the target space $S^{2}$. The discretized Euclidean action then takes the simple form
\begin{equation}
    S = \frac{1}{2 g_0^2} \sum_{I,x,\mu} a^2 \; \partial^f_\mu \phi^I(x) \partial^f_\mu \phi^I(x)
\end{equation}
where $\partial^f$ denotes the discretized forward derivative and the field satisfies the constraint $\sum_I \phi^I \phi^I = 1$. The energy momentum tensor can be obtained as the Noether current associated with translational invariance. In the continuum, it is given by
\begin{equation}
    T_{\mu\nu}(x) = \frac{1}{g_0^2} \sum_I \left[ \partial_{\mu} \phi^I(x) \partial_{\nu} \phi^I(x) - \frac{1}{2} \delta_{\mu\nu} \sum_\eta \partial_{\eta} \phi^I(x) \partial_{\eta} \phi^I(x)\right].
\end{equation}
Using Dyson-Schwinger relations as well as local Ward identities from the $O(3)$ symmetry \cite{Lauk:2025agt}, one can show that the renormalized energy momentum tensor on the lattice can be written as a linear combination of its decomposition into irreducible representations of the group of discretized Euclidean spacetime rotations $D_4$:
\begin{equation}
    \begin{aligned}
        T^1_{\mu\nu}(x) &= \frac{1}{g_0^2}\delta_{\mu\nu} \sum_{I,\rho} \partial^s_\rho \phi^I(x) \partial^s_\rho \phi^I(x) , \\
        T^2_{\mu\nu}(x) &= \frac{1}{g_0^2}\delta_{\mu\nu} \sum_I\left[ \partial^s_\mu \phi^I(x) \partial^s_\nu \phi^I(x) - \frac{1}{2} \sum_\rho \partial^s_\rho \phi^I(x) \partial^s_\rho \phi^I(x)\right] , \\
        T^3_{\mu\nu}(x) &= \frac{1}{g_0^2}(1 - \delta_{\mu\nu}) \sum_I \partial^s_\mu \phi^I(x) \partial^s_\nu \phi^I(x).
        \end{aligned}
\end{equation}
Here, $\partial^s$ denotes the discretized symmetric derivative. 
Since the theory is asymptotically free and the renormalization pattern is dictated by the short-distance regime, it is expected to hold also at the non-perturbative level. We calculate the renormalization constants non-perturbatively, focusing on the sector containing $T^2$ and $T^3$. With this in mind, we write the renormalized EMT as
\begin{equation}\label{renEMT}
    \left[T_{\mu\nu} (x) \right]_r = Z_T \left\{ z_s \left(T_{\mu\nu}^1 (x) - \langle T_{\mu\nu}^1 (x) \rangle \right) + z_T T_{\mu\nu}^2 (x) + T_{\mu\nu}^3 (x)\right\}.
\end{equation}

\subsection{The $g_0 \to 0$ limit}
To be able to perform a tree-level subtraction as well as provide a crosscheck for our numerical implementation we additionally consider the $g_0 \to 0$ limit of the theory. To this end, we want to perform a saddle point expansion. However, the partition function does not have a unique minimum around which to expand. To address this, we split off the integration over a single lattice site $x_0$, keeping it non-perturbative. The integral over the rest of the points now has a unique minimum and a saddle point approximation is possible. We use the parametrization
\begin{equation}
    \phi(x) = \frac{\phi(x_0) + g_0 z(x)}{||\phi(x_0) + g_0 z(x)||}
\end{equation}
for which the path integral for some operator $P$ becomes
\begin{equation}
    \langle P \rangle = \frac{1}{Z} g_0^{2(N_0N_1-1)}\int d\Omega(\phi(x_0)) \int \prod_{x\neq x_0} d^{2} z(x) \;e^{-S(\phi(z))} P(\phi(z)) + \mathcal{O}(g_0^2).
\end{equation}
The right hand side, after additionally expanding action and operator in $g_0$, becomes a sequence of gaussian integrals. After evaluating the integrals for $P = \phi^I(x) \phi^I(y) -  1$, we obtain
\begin{equation}
    \left\langle \sum_I \phi^I(x) \phi^I(y) -  1 \right\rangle = g_0^2 \left(J_{xx} + J_{yy} - 2J_{xy}\right) + O(g_0^3)
\end{equation}
where $J$ is the inverse of the lattice Laplacian 
$\sum_{\mu}[\delta_{x-\hat\mu,y} + \delta_{x+\hat\mu,y}-2\delta_{xy}]$ 
restricted to $x, y \neq x_0$, and $J_{x_0, y} = J_{x, x_0} = 0$. Although $J$ itself depends on the choice of $x_0$, the combination $J_{xx} + J_{yy} - 2J_{xy}$ is translationally invariant. The matrix $J$ cannot be obtained in closed form, but its elements can be computed numerically using a standard minimization procedure with sources.

\subsection{Lattice setup}
To generate configurations we use a Wolff cluster algorithm \cite{Wolff:1988uh}. Since this algorithm is highly efficient and we obtain many decorrelated configurations (up to $\mathcal{O}(10^9)$ per run), we calculate the observable mean on the fly using improved estimators \cite{Wolff:1989hv} and discard the configurations as well as individual measurements. To obtain robust errors, we calculate the autocorrelation function on the fly up to a maximum lag and then extract the error estimates using the gamma method.

Both the generation and analysis code can be made available upon reasonable request.

\section{Choice of action and renormalization scheme}\label{cutofftheo}
$O(N)$ nonlinear sigma models on the lattice exhibit large discretization artifacts,  which can be understood in a Symanzik EFT framework via the appearance of operators with large anomalous dimensions (large logarithms) \cite{Balog:2009np}. One approach that has been used successfully to reduce discretization artifacts for specific observables, though it does not correspond to Symanzik improvement, is to use the constrained action \cite{Balog:2012db}
\begin{equation}
    \begin{aligned}
        S_{\text {oca }}[\phi]=
        \begin{cases}
        S[\phi] \qquad \text{ for } \sum_I \phi(x)^I \phi(x+\hat{\mu})^I > \cos(\delta)\\
        \infty \qquad \text{ else },
        \end{cases}
    \end{aligned}
\end{equation}
restricting the angle between neighboring spins to not exceed a maximal value $\delta$. While in ref. \cite{Balog:2012db} an optimal reduction of discretization artifacts was found for $\cos(\delta) = -0.345$, we choose to make the constraint dependent on the bare coupling in a linear fashion, setting
\begin{equation}
    \cos(\delta) = 1-1.345g_0.
\end{equation}
We compare all three actions and their approach to the $g_0 \to 0$ limit in Section ~\ref{cutoffeffsec}.

\medskip
When using the constrained action, the Gaussian integrals in the saddle point expansion run over a region bounded by the constraint rather than all of $\mathbb{R}^{2}$. However, one can show that for a dependence of the form $\cos(\delta) = 1-cg_0$, the saddle point approximation is valid up to corrections of order $\exp(-c/g_0)$ which can thus be safely disregarded. 

\medskip
Another important issue that the constraint introduces is that of topological freezing. For values of $\delta < \pi/2$, the energy barriers between topological sectors become infinite. To avoid issues with ergodicity, even at $\delta > \pi/2$, we restrict all measurements to the trivial topological sector $Q=0$. We employ the gradient flow \cite{Makino:2014sta}, via the flowtime equation
\begin{equation}
    \partial_t \phi^I(x, t) = \sum_J \left [ (\delta^{IJ} - \phi^I(x, t)\phi^J(x, t)) \Box \phi^J(x, t)\right].
\end{equation}
Similarly to the case of Yang-Mills theory, $O(3)$ invariant operators built from flowed fields are renormalized quantities for positive flowtimes $t > 0$. This allows us to define the renormalized coupling at scale $\mu = 1/(0.6L_0)$ and topological charge $Q=0$ (similar to \cite{Fritzsch:2013yxa}), via
\begin{equation}
    g_{GF}^2\left(\mu = 1/\sqrt{4t}\right) \coloneqq \mathcal{N}^{-1} \frac{t\langle E(t) \delta_Q \rangle}{\langle \delta_Q \rangle}\Bigg|_{\sqrt{4t} = 0.6L_0}
\end{equation}
where $E(t) = g_0^2 S(t)/V$ is the action density constructed from flowed fields and $t$ is the flowtime. Throughout this work, we also choose to fix $L_1 = 10L_0$. The gradient flow is implemented using a standard $3rd$ order Runge-Kutta integrator.

\subsection{Choice of action}\label{cutoffeffsec}

To motivate our choice of action, we compare the gradient flow coupling $g^2_{\mathrm{GF}}$ as a function of bare coupling for three different actions: the standard (unconstrained) action, the optimized constraint action with $\cos(\delta) = -0.345$ from ref. \cite{Balog:2012db}, and our modified constraint action with $\cos(\delta) = 1 - 1.345g_0$. We perform this comparison at fixed $N_0 = 6, 12, 18$, allowing us to study different lattice spacings at the same physical volume.

\begin{figure}
    \centering
    \begin{subfigure}{0.49\columnwidth}
        \includegraphics[width=\linewidth]{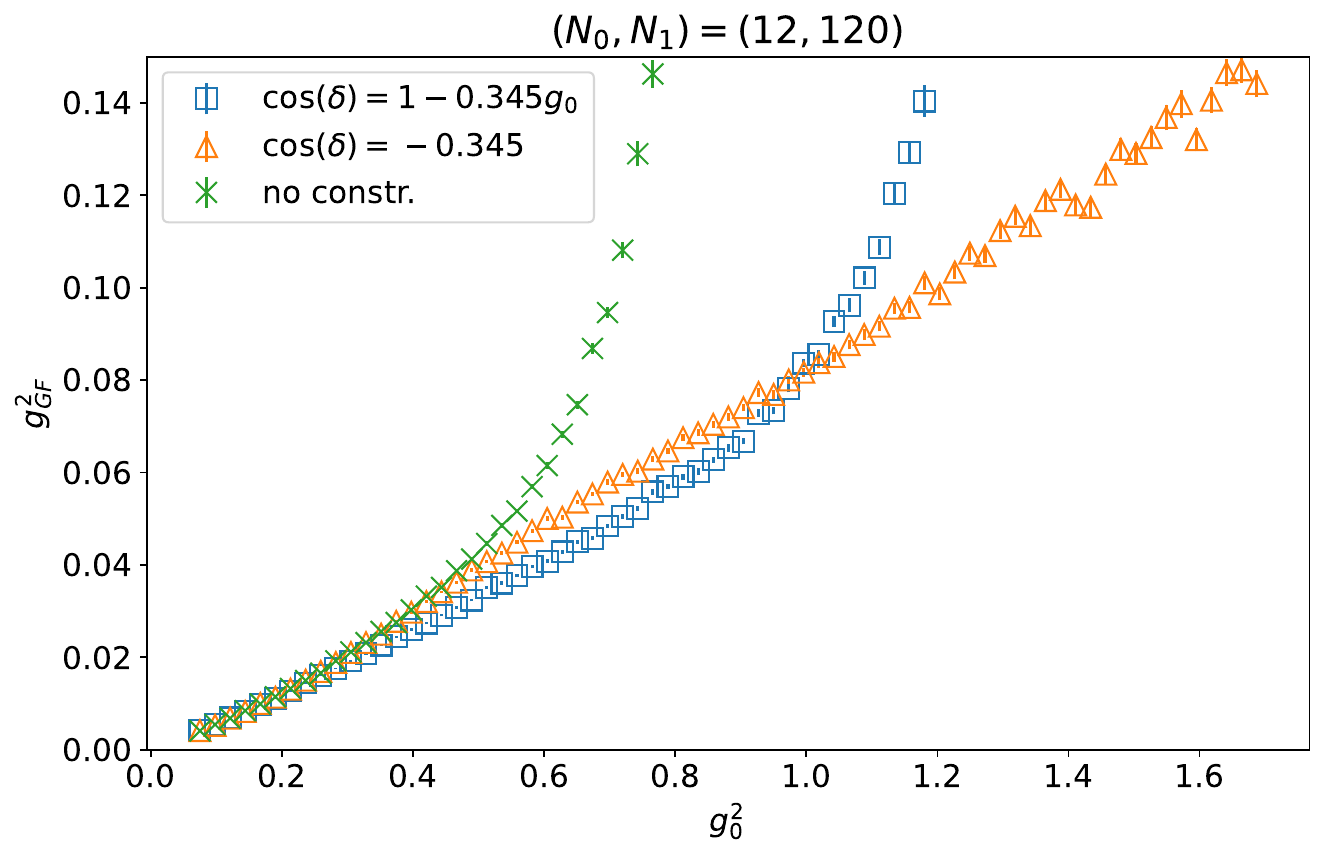}
        \caption{}
        \label{fig:sub1}
    \end{subfigure}
    \begin{subfigure}{0.49\columnwidth}
        \includegraphics[width=\linewidth]{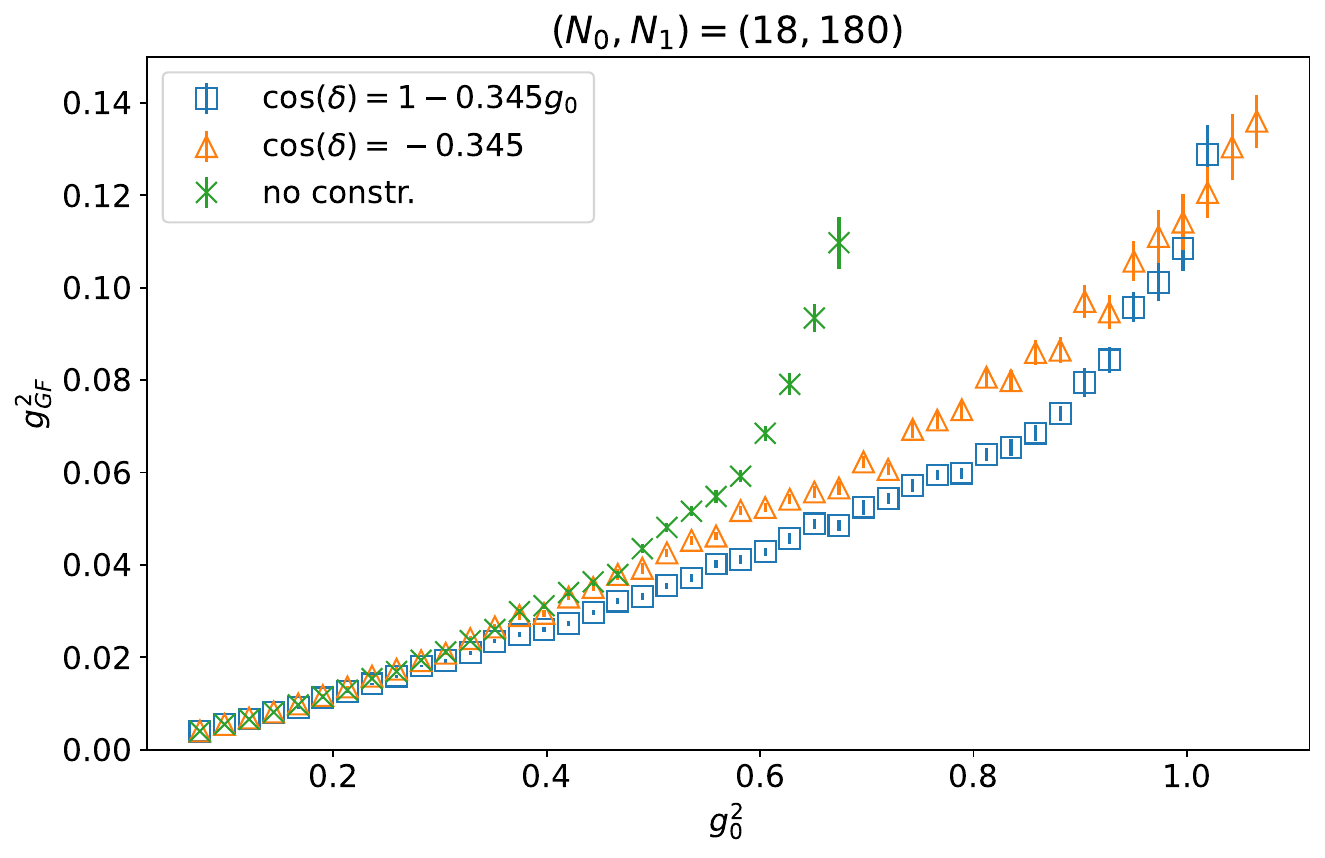}
        \caption{}
        \label{fig:sub2}
    \end{subfigure}
    \caption{\label{cutoff_eff} Gradient flow coupling $g^2_{\mathrm{GF}}$ as a function of bare coupling $g^2_0$ for three different actions at $N_0 = 12$ (a) and $N_0 = 18$ (b). For renormalized couplings below $g^2_{\mathrm{GF}} \approx 0.08$, the modified constraint action reaches a given value of $g^2_{\mathrm{GF}}$ at larger bare coupling.}
    \label{fig:both}
\end{figure}

The results for $N_0 = 12$ and $18$ are shown in Figure \ref{cutoff_eff} and are representative of the case $N_0 = 6$. For renormalized couplings below $g^2_{\mathrm{GF}} \approx 0.08$, the modified constraint action reaches a given value of $g^2_{\mathrm{GF}}$ at larger bare coupling than the other actions. At the same bare coupling, using the modified constraint action thus corresponds to simulating a finer lattice. We adopt the modified constraint action for all subsequent measurements, but whether this also reduces discretization artifacts in the observables of interest is not guaranteed.

\begin{figure}[H]
    \centering
    \begin{subfigure}{0.49\columnwidth}
        \includegraphics[width=\linewidth]{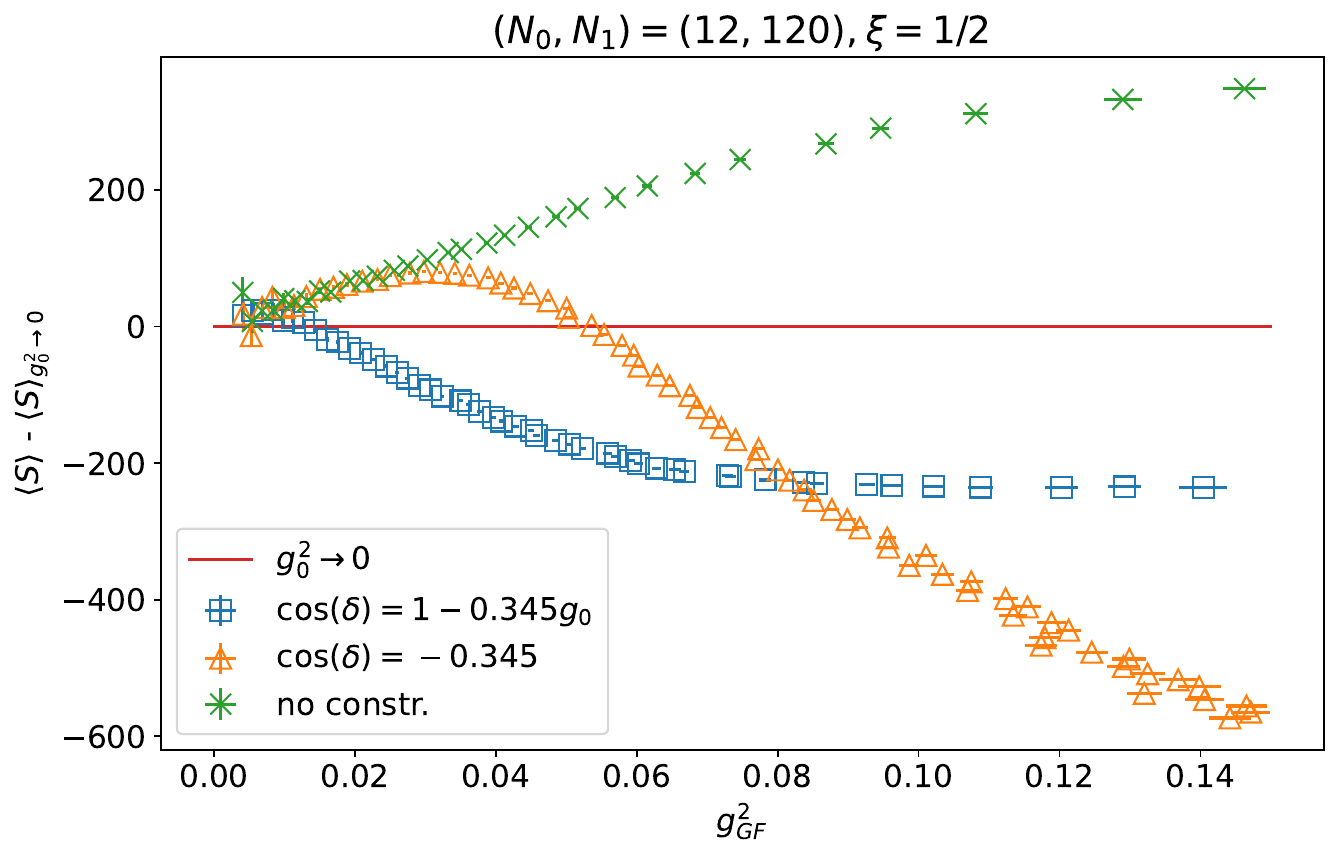}
        \caption{}
        \label{fig:sub1}
    \end{subfigure}
    \begin{subfigure}{0.49\columnwidth}
        \includegraphics[width=\linewidth]{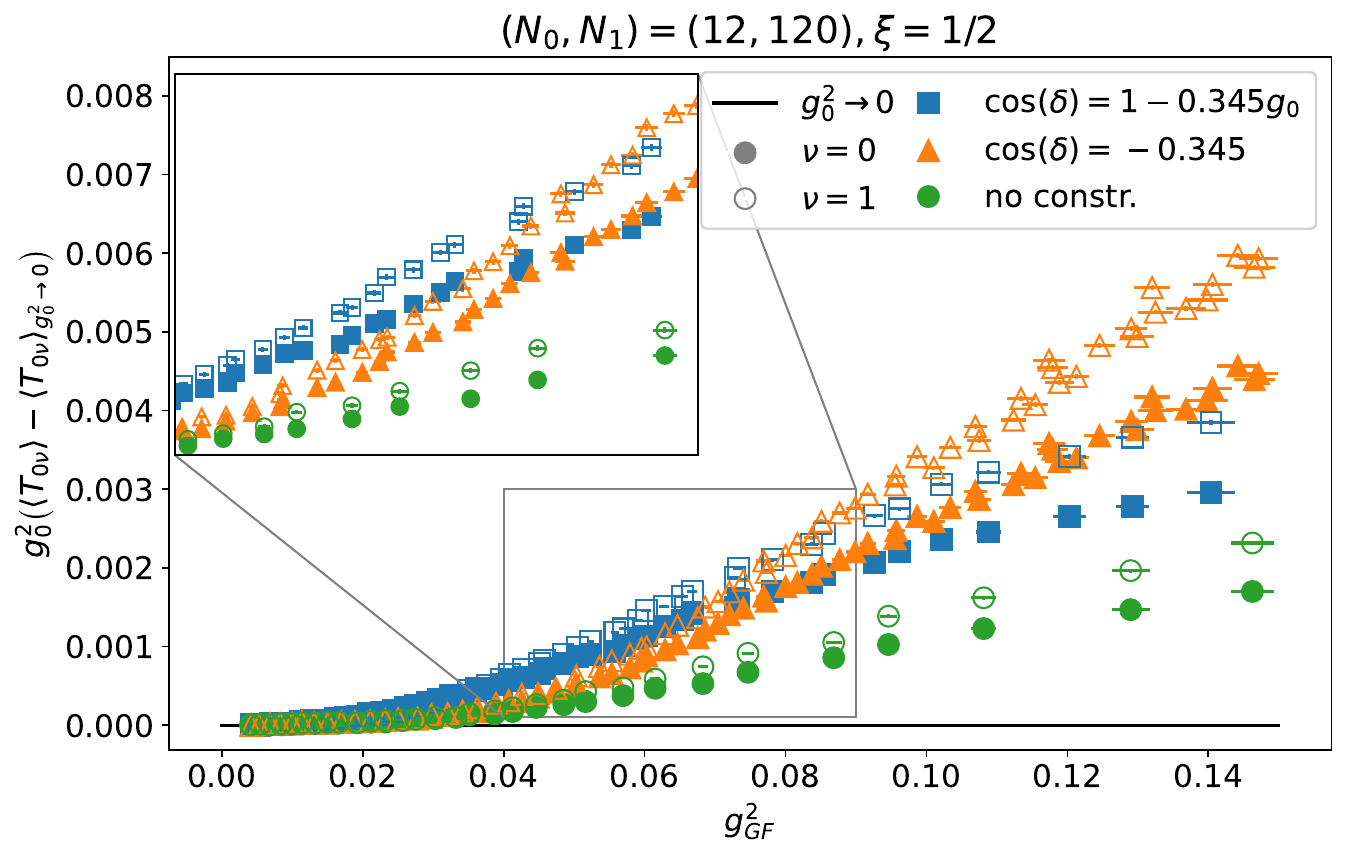}
        \caption{}
        \label{fig:sub2}
    \end{subfigure}
    \caption{\label{cutoff_eff_obs}$g_0 \to 0$ approach for the action $\langle S \rangle - S_{g^2_0 \to 0}$ (a) and the EMT one-point functions $\langle T_{0\nu} \rangle - \langle T_{0\nu} \rangle_{g^2_0 \to 0}$ (b) as a function of $g^2_{\mathrm{GF}}$ at $N_0 = 12$ and $\xi = 1/2$. The action density shows qualitatively different behaviour from the EMT one-point functions, while the latter show similar deviations from the free-theory limit across all three actions.}
    \label{fig:both}
\end{figure}

To assess how different observables behave across the three actions, we also calculate the action $\langle S \rangle$ and the EMT one-point functions $\langle T_{0\nu} \rangle$ as functions of $g^2_{\mathrm{GF}}$ for $N_0 = 6,12,18$ with shifted boundary conditions in time (see Section \ref{renconstsec}). The results for $N = 12$ are shown in Figure \ref{cutoff_eff_obs} and are representative of the other cases. All quantities are shown relative to their $g^2_0 \to 0$ limit. 

In general, this is not a measure of the discretization artifacts. What is notable, however, is that $\langle T_{00} \rangle$ and $\langle T_{01} \rangle$ show a strikingly similar dependence on $g^2_{\mathrm{GF}}$ across all three actions and both values of $\nu$. In contrast, the action shows a qualitatively different dependence.

\FloatBarrier

\subsection{Tuning}\label{Tuningsec}
\begin{figure}[H]
    \centering
    \begin{subfigure}{0.48\columnwidth}
        \includegraphics[width=\linewidth]{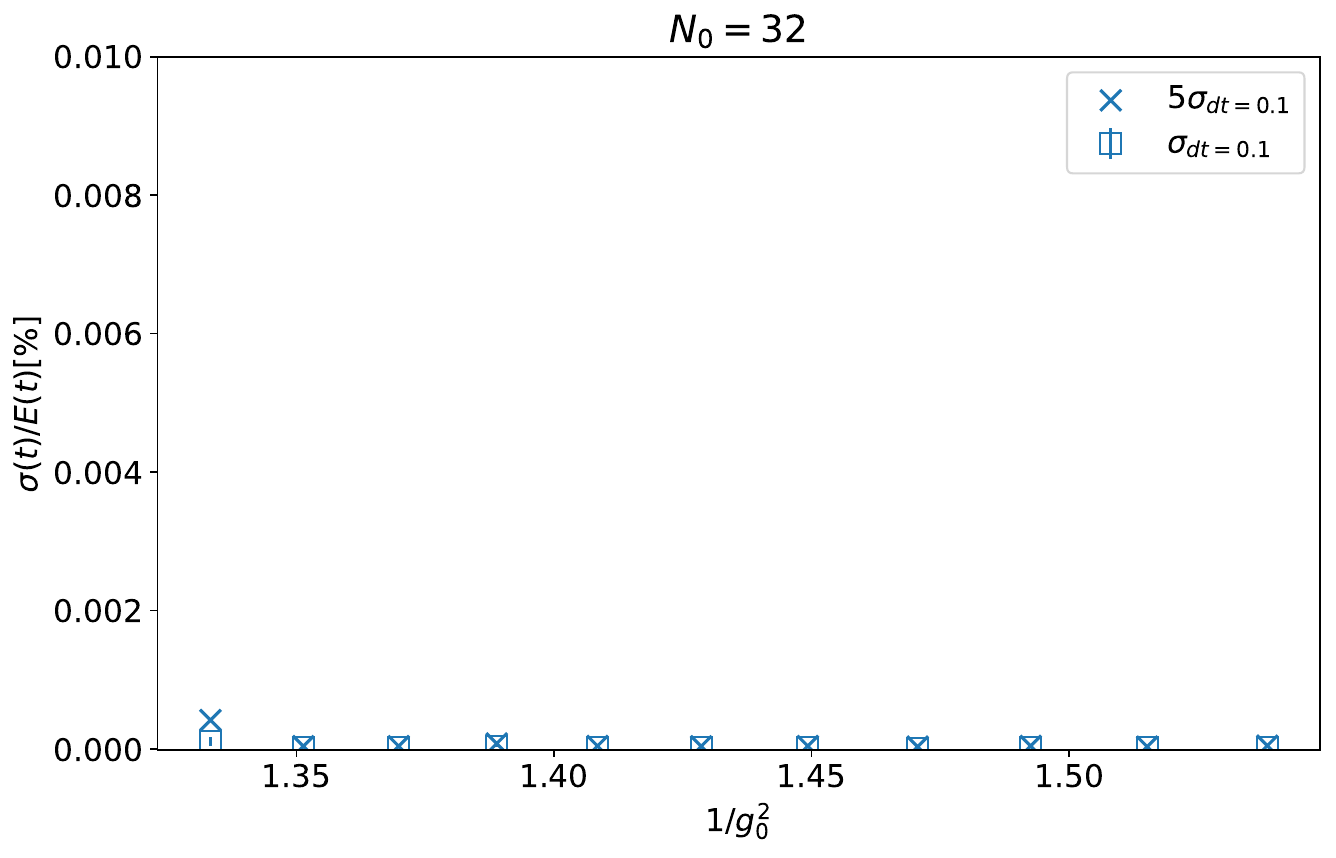}
        \caption{}
        \label{figtune:sub1}
    \end{subfigure}
    \begin{subfigure}{0.485\columnwidth}
        \includegraphics[width=\linewidth]{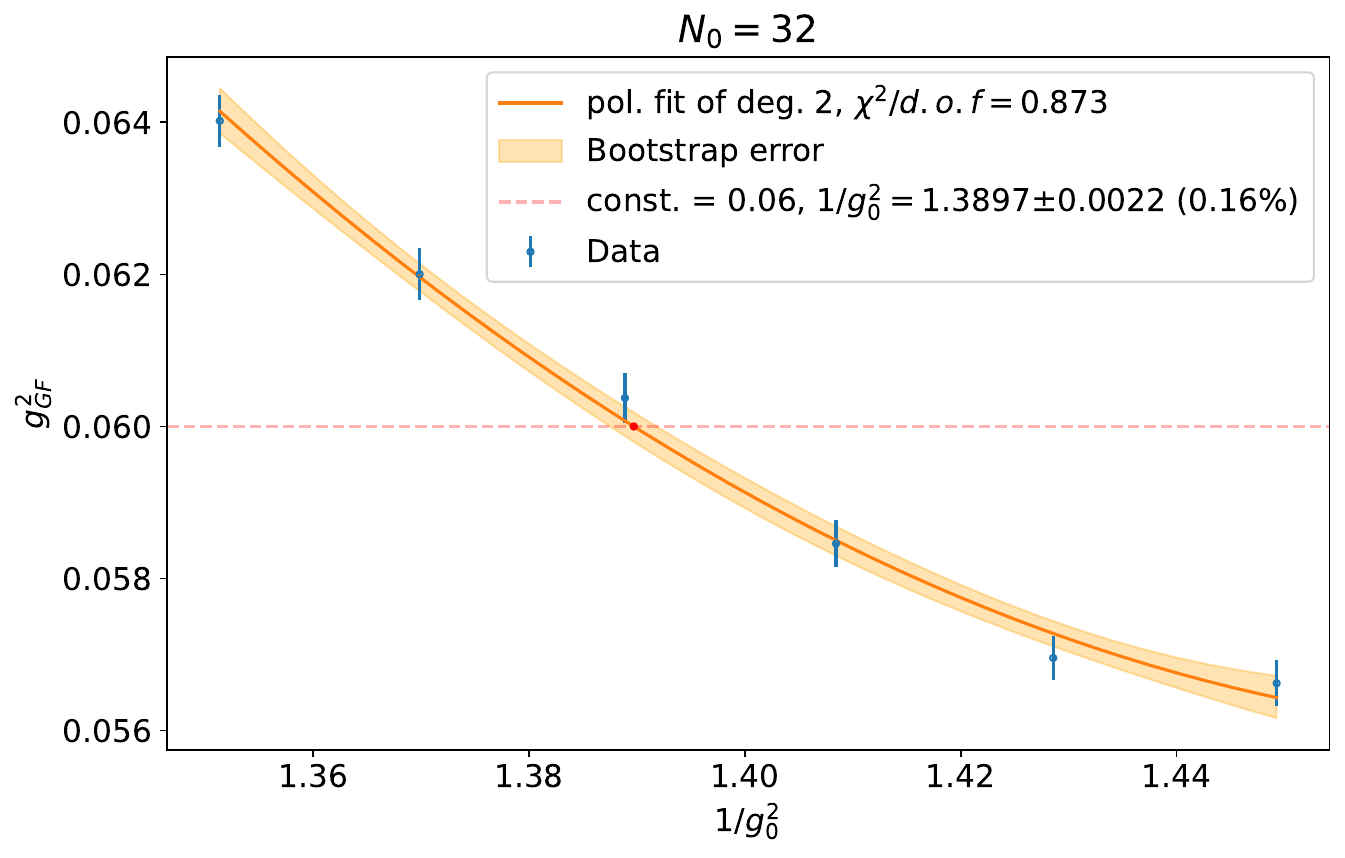}
        \caption{}
        \label{figtune:sub2}
    \end{subfigure}
    \caption{\label{tuningfig}Systematic effects from flowtime discretization (a) and tuning of the bare coupling (b) for $N_0 = 32$. In (a), we show that discretization artifacts are well below $0.01\%$ for our chosen step size. In (b), we fit a second order polynomial to determine the bare coupling at $g^2_{\mathrm{GF}} = 0.06$, with errors estimated via bootstrap.}
    \label{fig:both}
\end{figure}

We determine $1/g_0^2$ at $N_0 = 6, 8, 10, 12, 18, 32$. Our line of constant physics is chosen as $g_{GF}^2 = 0.06$. In this case we find $1/m_{eff} \sim 2L_0 \ll L_1 = 10L_0$, so that finite-size effects from the spatial extent are negligible, while sensitivity to the finite temporal extent remains. We verify that the systematic effects stemming from the flowtime discretization are below $0.01\%$ by at least $5\sigma$ (see Figure \ref{figtune:sub1}). Once this is ensured, we fit a second order polynomial to the data and find the intersection with the chosen value of our gradient-flow coupling (see Figure \ref{figtune:sub2}). The error of the intersection is estimated by a bootstrap procedure on the fit. The fit range is chosen by a quality of fit criterion, although we found that determinations from all ranges are compatible within $1\sigma$ of each other. The results, including their $\chi^2/d.o.f$ values can be found in Table \ref{tuningvals}.

\begin{table}[h!]
    \centering
    {$
    \begin{array}{c|c|c|c|c|c|c}
         N_0 & 6 & 8 & 10 & 12 & 18 & 32\\ \hline
         1/g_0^2 & 1.0939(10) & 1.1405(21) & 1.1760(27) & 1.2091(28) & 1.2884(27) & 1.3897(22)\\ \hline
         \chi^2/d.o.f& 0.839 & 1.015 & 0.436 & 0.88 & 1.205 & 0.873\\
    \end{array}
    $}
    \caption{\label{tuningvals} Tuned values of $1/g^2_0$ along the line of constant physics $g^2_{\mathrm{GF}} = 0.06$ for various temporal extents $N_0$, including the $\chi^2/d.o.f$ of the fits.}
\end{table}

\FloatBarrier

\section{Calculation of renormalization constants}\label{renconstsec}
To extract the renormalization constants, we use shifted boundary conditions \cite{Giusti:2012yj}, a method that has been previously used successfully in renormalizing the EMT in Yang-Mills theory \cite{Giusti:2015daa}. Here one considers the thermal theory in a moving frame, the finite temperature partition function being
\begin{equation} \label{partfuncfinite}
		Z = \text{Tr}(\exp(-L_0(H - i\xi P)))
\end{equation}
where $H$ is the Hamiltonian and $P$ the momentum operator. In a finite volume $ V = L_0 L_1$ this corresponds to a euclidean path integral with shifted boundary conditions in time, i.e.
\begin{equation}
    \begin{aligned}
        &\phi(L_0, x_1) = \phi(0, x_1 - \xi L_0) \\
        &\phi(x_0, L_1) = \phi(x_0, 0).
    \end{aligned}
\end{equation}
In the continuum, the partition function \eqref{partfuncfinite} has an $SO(2) \times SL(2,\mathbb{Z})$ symmetry, which gives rise to Ward identities relating different components of the energy-momentum tensor. The relevant identities are \cite{Giusti:2012yj}
\begin{align}
    \langle T_{01} \rangle_\xi &= \frac{1}{L_0 L_1} \frac{\partial}{\partial \xi} \ln Z(L_0, L_1, \xi), \label{eq:ZT_log_cont} \\
    \frac{\partial}{\partial \xi}\langle T_{01} \rangle_\xi &= L_0 \langle T_{01}(0,y_1) \sum_{x_1} T_{01}(x_0,x_1) \rangle_{\xi,c} \quad (x_0 \neq 0),  \label{eq:ZT_2p_cont} \\
    \langle T_{00} \rangle_\xi &= \frac{1-\xi^2}{2\xi} \langle T_{01} \rangle_\xi \quad \label{eq:zT_cont}
\end{align}
where the last identity holds when $L_1 \xi / ((1+\xi^2)L_0) \in \mathbb{Z}$. The subscript $\xi$ denotes that the expectation value is calculated at a specific shift $\xi$.
\medskip

The renormalization constants $Z_T$ and $z_T$ are defined within our finite-volume scheme at renormalization scale $\mu = 1/(0.6\,L_0)$, the physical volume being fixed by requiring $g^2_\mathrm{GF} = 0.06$ along the line of constant physics described in Section \ref{Tuningsec}. We determine $Z_T$ and $z_T$ by imposing the continuum Ward identities \eqref{eq:ZT_log_cont}-\eqref{eq:zT_cont} on the lattice, which fixes them up to discretization artifacts that vanish in the continuum limit. Along the line of constant physics, $g_0$ is fixed as a function of $N_0$ by the tuning in Table \ref{tuningvals}, so that $Z_T$ and $z_T$ are functions of a single parameter, which we may take to be either $g_0$ or $N_0$.

Concretely, we insert the lattice EMT \eqref{renEMT} into equations \eqref{eq:ZT_log_cont}-\eqref{eq:zT_cont}. We use a symmetric derivative at $\xi = 1/2$ for \eqref{eq:ZT_log_cont} and a forward derivative at $\xi = 0$ for \eqref{eq:ZT_2p_cont}, exploiting $\langle T_{01}\rangle_{\xi=0} = 0$ by rotational symmetry. This yields
\begin{align}
    Z_{T,\mathrm{log}}(g_0,N_0) &= \frac{1}{2aL_1} \frac{1}{\langle T_{01} \rangle_{1/2}} 
    \ln\!\left(\frac{Z(L_0, L_1, 1/2 + a/L_0)}{Z(L_0, L_1, 1/2 - a/L_0)}\right), \label{eq:ZT_log_lat} \\
    Z_{T,2p}(g_0,N_0) &= \frac{\langle T_{01} \rangle_{1/L_0}}{\langle T_{01}(0,y_1) \sum_{x_1} T_{01}(x_0,x_1) \rangle}, \quad x_0 \neq 0, \label{eq:ZT_2p_lat} \\
    z_T(g_0,N_0) &= \frac{3\langle T_{01} \rangle_{1/2}}{4\langle T_{00} \rangle_{1/2}}. \label{eq:zT_lat}
\end{align}
The subscripts $log$ and $2p$ in equations \eqref{eq:ZT_log_lat}, \eqref{eq:ZT_2p_lat} emphasize that we use two different Ward identities to determine the same renormalization constant $Z_T$. These determinations in principle differ in their subleading discretization artifacts.
The ratio of partition functions on the right hand side in \eqref{eq:ZT_log_lat} can be evaluated in multiple ways. One approach \cite{Giusti:2015daa} exploits the fact that the ratio is a smooth function of $g^2_0$ and integrates its logarithmic derivative over a range of bare couplings. However, we found this to be computationally infeasible at the required precision. Our approach is to use simulated tempering \cite{Marinari:1992qd} with respect to the shift $\xi$, treating it as a free parameter in the simulation that is updated with a Metropolis accept-reject step.

\medskip
These expressions should in principle be evaluated at fixed topological charge $Q = 0$, consistently with the definition of the gradient flow coupling.
However, projecting to $Q = 0$ would require integrating the gradient flow equation on every configuration, which is computationally prohibitive at the required statistics. We therefore evaluate these identities without explicit projection to $Q = 0$, since the vast majority of configurations in our ensembles already lie in this sector. This possibly introduces a systematic uncertainty that remains to be quantified.

\subsection{Numerical results}

\begin{figure}
    \centering
    \begin{subfigure}{0.48\columnwidth}
        \includegraphics[width=\linewidth]{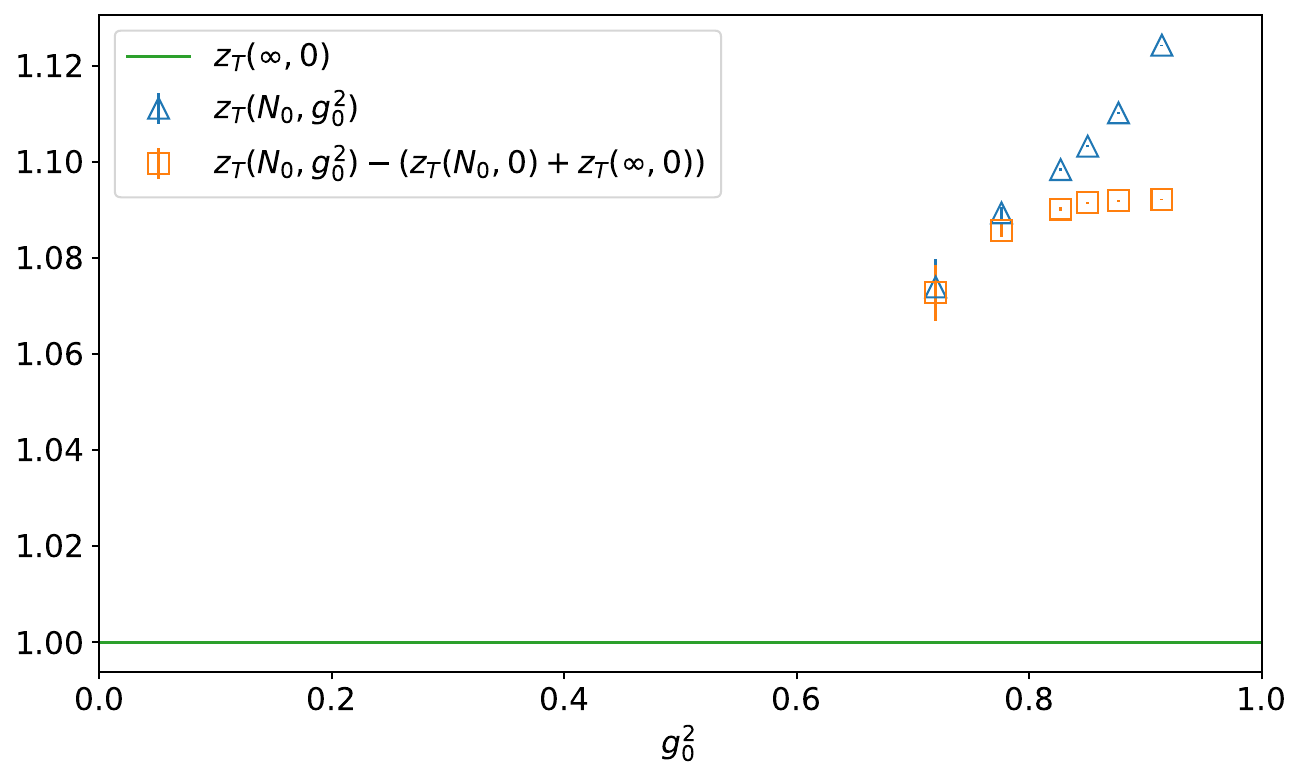}
        \caption{}
        \label{fig:sub1}
    \end{subfigure}
    \begin{subfigure}{0.49\columnwidth}
        \includegraphics[width=\linewidth]{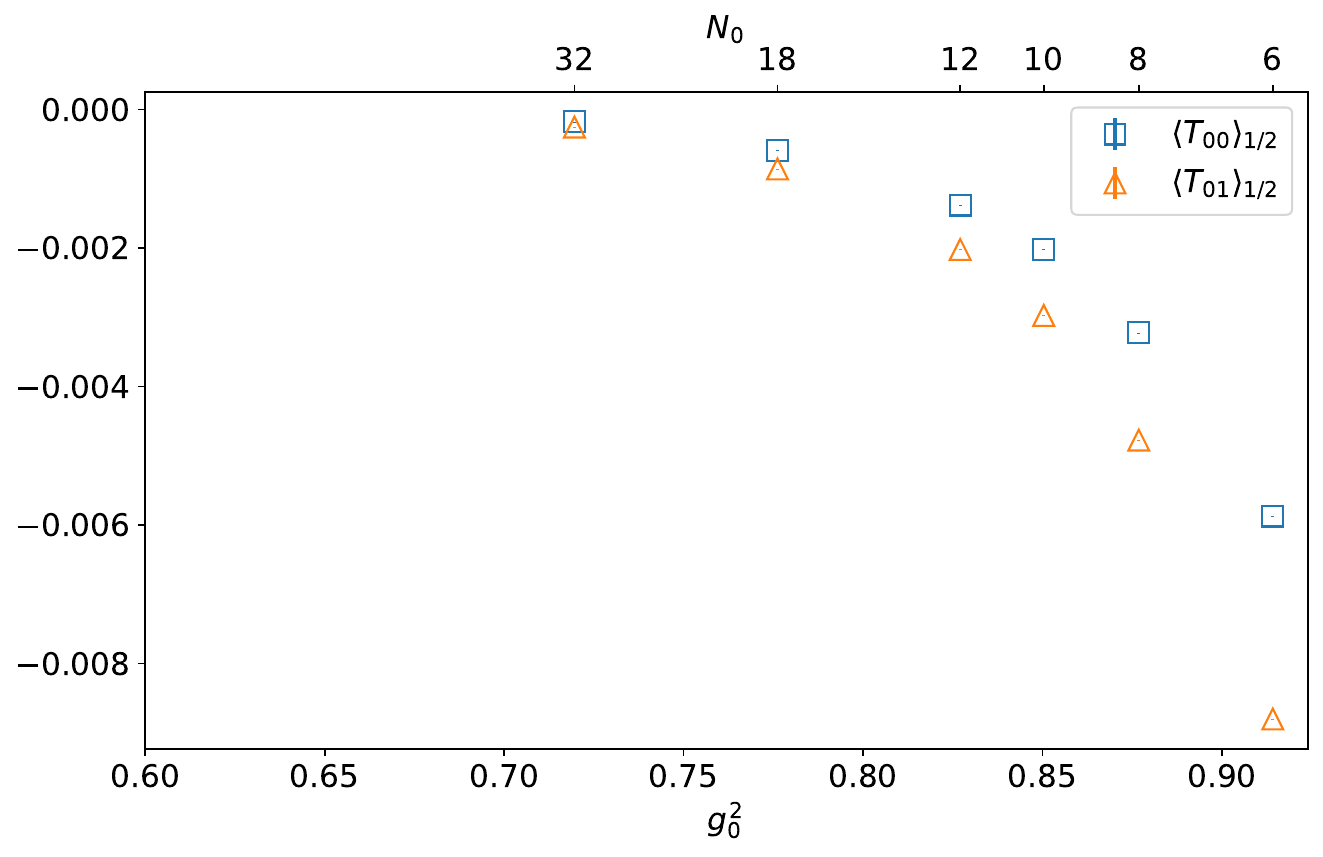}
        \caption{}
        \label{fig:sub2}
    \end{subfigure}
    \caption{\label{resultszT} Results for $z_T$ as a function of $g^2_0$ (a) and the individual EMT one-point functions entering its determination (b). The tree-level subtraction worsens results for $N_0 \leq 18$ and makes no difference for $N_0 = 32$. The similar deviations of $\langle T_{00}\rangle_{1/2}$ and $\langle T_{01}\rangle_{1/2}$ from their free-theory values may lead to cancellations in the ratio defining $z_T$.}
    \label{fig:both}
\end{figure}

We calculate $z_T$ using equation \eqref{eq:zT_lat}. The results, with and without tree-level subtraction, are shown in Figure \eqref{resultszT} and achieve sub-percent precision. Here, we benefit from the fact that both numerator and denominator are evaluated at the same shift. Thus we can calculate them on the same ensemble and have cancellations in the autocorrelation function. Additionally, as discussed in Section \ref{cutoffeffsec}, both numerator and denominator deviate from their $g_0 \to 0$ limits in a similar way, possibly leading to partial cancellations in the ratio. Interestingly, the tree-level subtraction worsens the results for $N_0 \leq 18$ while making no discernible difference for $N_0 = 32$, suggesting that the $g_0 \to 0$ limit provides a poor approximation at the coarser lattice spacings.

\begin{figure}
    \centering
    \begin{subfigure}{0.47\columnwidth}
        \includegraphics[width=\linewidth]{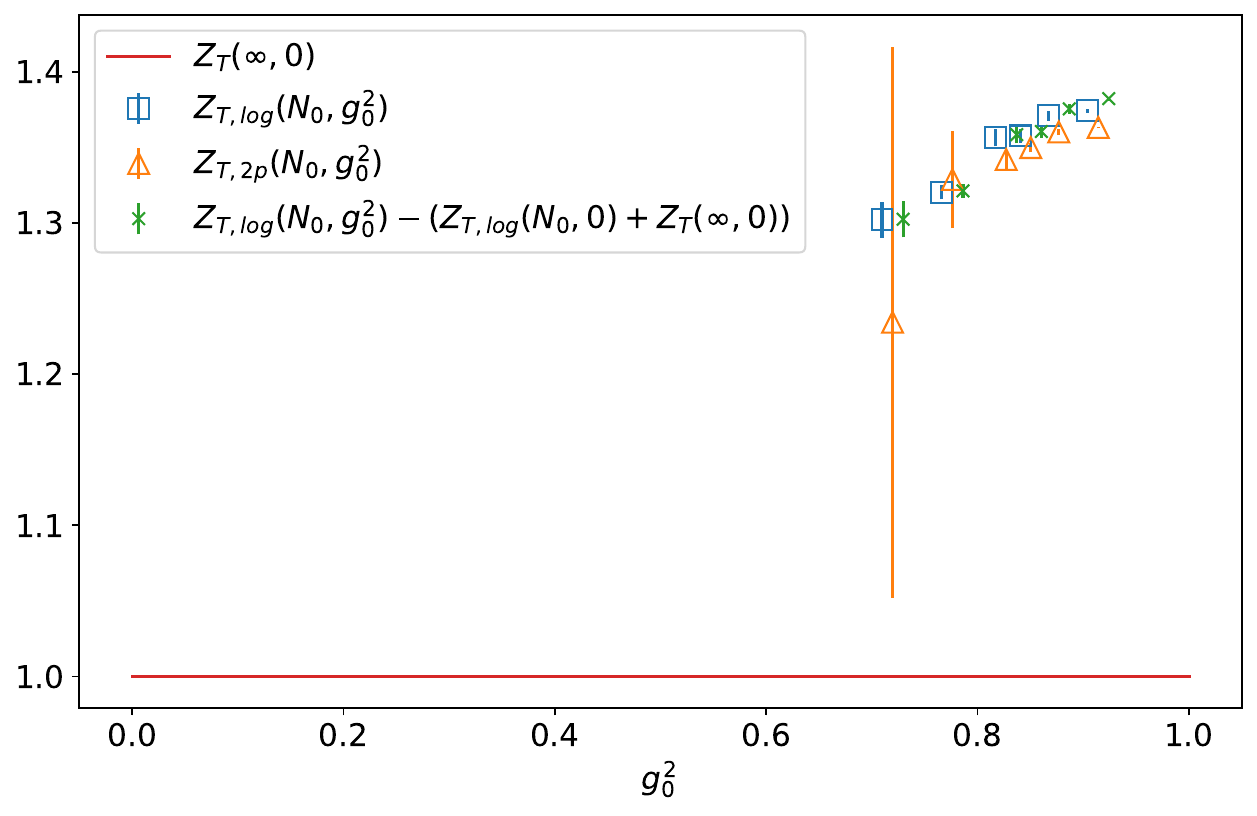}
        \caption{}
        \label{fig:sub1}
    \end{subfigure}
    \begin{subfigure}{0.52\columnwidth}
        \includegraphics[width=\linewidth]{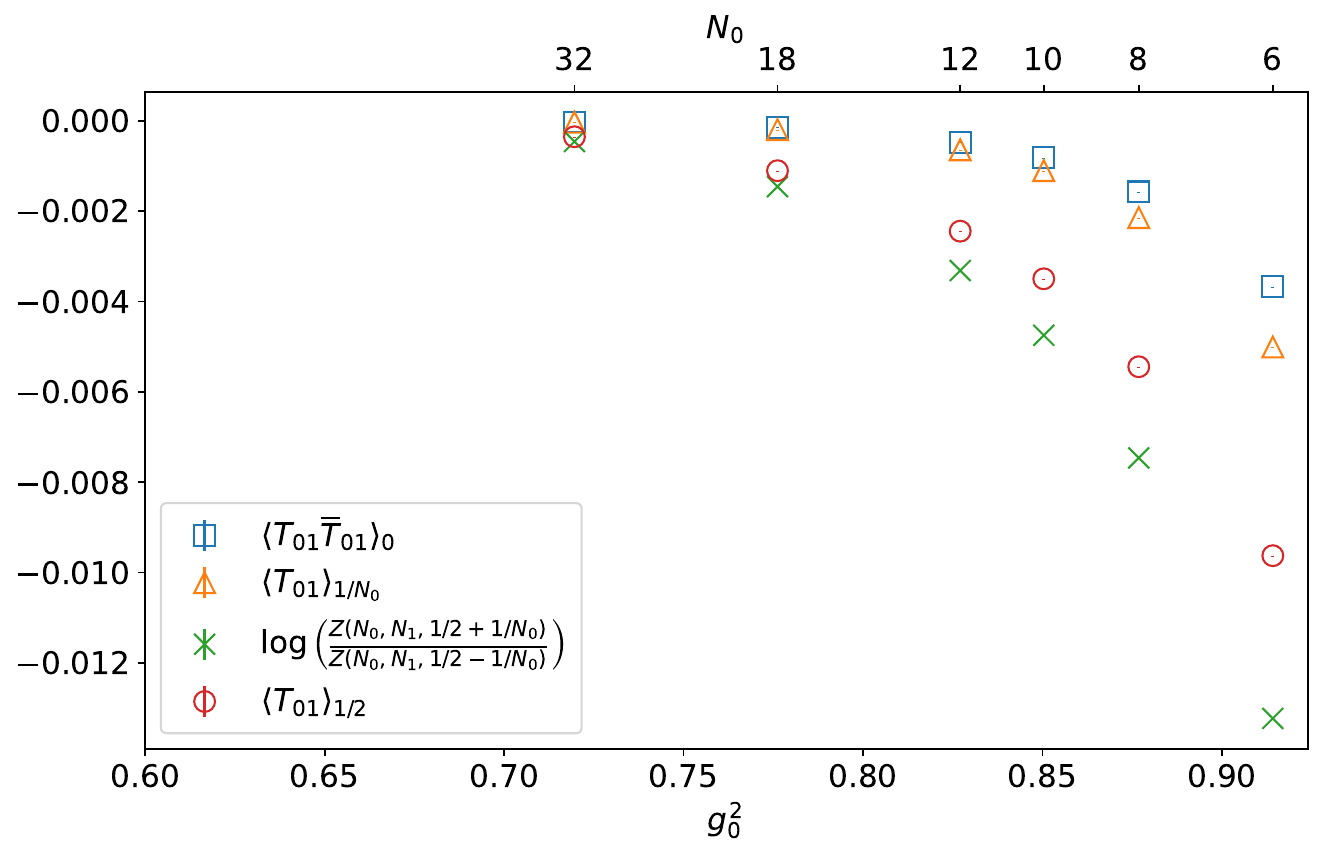}
        \caption{}
        \label{fig:sub2}
    \end{subfigure}
    \caption{\label{resultsZT} Results for $Z_T$ as a function of $g^2_0$ (a) and the individual observables entering its determination (b). In (a), the points have been offset horizontally for better visibility. Both methods, $Z_{T,\mathrm{log}}$ from equation \eqref{eq:ZT_log_lat} and $Z_{T,2p}$ from equation \eqref{eq:ZT_2p_lat} are mutually compatible, indicating that the dominant discretization artifacts are common to both.}
    \label{fig:both}
\end{figure}

We now turn to $Z_T$, evaluated using equations \eqref{eq:ZT_log_lat} and \eqref{eq:ZT_2p_lat}. The results are shown in Figure \ref{resultsZT}. Unlike in the determination of $z_T$, the two observables entering $Z_{T,\mathrm{log}}$ and $Z_{T,2p}$ are evaluated at different shifts, so there are no cancellations in the autocorrelation function, making the determination of $Z_{T,2p}$ especially hard. The results allow for a continuum extrapolation and are mutually compatible between both methods. Since the two methods have different subleading discretization artifacts, their agreement strongly suggests that the dominant $O(a^2)$ artifacts are common to both.

\section{Conclusion and Outlook}

We have presented a non-perturbative study of the energy-momentum tensor renormalization in the two-dimensional $O(3)$ nonlinear sigma model using shifted boundary conditions and a gradient flow scheme at fixed topological charge $Q = 0$. After comparing three different lattice actions, we adopted a modified constraint action with $\cos(\delta) = 1 - 1.345g_0$.
 
For the mixing constant $z_T$, we obtain results with sub-percent precision. Here, we benefit from the fact that both observables entering the ratio are evaluated at the same shift, allowing for cancellations in the autocorrelation function. Additionally, the similar deviations of $\langle T_{00}\rangle$ and $\langle T_{01}\rangle$ from their $g_0 \to 0$ approximation may lead to partial cancellations in the ratio.
 
The determination of $Z_T$ is also possible with sub-percent precision, provided one uses the tempering approach and calculates the ratio of partition functions directly. Both methods we employ are mutually compatible, which indicates that the dominant $O(a^2)$ discretization artifacts are common to both.

\medskip
Several directions for future work present themselves. Firstly, we need to quantify the possible systematic error introduced by not projecting the observables themselves to $Q = 0$. Afterwards, one possibility is to explore alternative ways to determine the renormalization constants such as Ward identities at positive flowtime \cite{DelDebbio:2013zaa} or a small-flowtime expansion \cite{Makino:2014sta}. However, since the dominant discretization artifacts appear to originate from operators with large anomalous dimensions in the Symanzik expansion and are a property of the lattice theory itself, alternative extraction methods might not yield significantly better results. Another way forward is to investigate whether Symanzik improvement could reduce them. However, a Symanzik-improved action would not be straightforwardly amenable to the Wolff cluster algorithm in the sense that it would lead to critical slowing down. It remains to be seen whether the reduction of discretization artifacts is large enough to warrant using this improved action.

\acknowledgments
The research of M.L. is funded by the Deutsche Forschungsgemeinschaft (DFG, German Research
Foundation) - Projektnummer 417533893/GRK2575 ”Rethinking Quantum Field Theory”.

\bibliographystyle{JHEP}
\bibliography{main.bib}

\end{document}